\documentclass[useAMS,usenatbib,usegraphicx]{mn2e}

\usepackage{amsmath}
\title{Magnetohydrodynamic instability in differentially rotating compressible flow\\}
\author[Mradul Sharma]{Mradul Sharma\thanks{E-mail:
mradul@barc.gov.in}\\
Astrophysical Sciences Division, Bhabha Atomic Research Centre, Mumbai - 400085, India\\}
\begin{document}



\maketitle

\label{firstpage}

\begin{abstract}
Transport of angular momentum is one of the thrust areas of astrophysical flows. Instabilities and hence the turbulence generated by it has been invoked to understand its role in angular momentum transport in hydrodynamic and magnetohydrodynamic regime. Investigation of an unexplored region described by the parameter space  $\Omega_e^2 < 0$, $\Omega_e$ being the epicyclic frequency, resulted in a powerful magnetohydrodynamic instability. The growth rate of this instability is rather large and is found to depend on the wavenumber. 
\end{abstract}


\section{Introduction}
Transport of angular momentum in various astrophysical flows is of great importance (\cite{Lyn,Shakura}) in accretion theory. Nature of accretion process is largely understood through theoretical and computational studies. In astrophysical flows, outward transport of angular momentum of accreting material is the basic requirement to sustain the accretion disk. This angular momentum transport can be sustained by some turbulent mechanism operating in accretion disk. The origin of turbulence is unclear. Turbulent viscosity ($\alpha$-viscosity) model by Shakura and Sunayev (\cite{Shakura}) is one of such mechanisms.
Differential rotation and the presence of magnetic field plays an important role in the dynamics of instability and hence in efficient angular momentum transport. In early $1990$s, Balbus \& Hawley (\cite{Balbus}) found that the subthermal fields in the presence of differential rotation can cause Magneto-Rotational Instability (MRI), which can provide the physical basis for the transport of angular momentum in accretion disk. Although the MRI requires only a small amount of ionization to work, it cease to exist when the strength of magnetic field increases. 
Paper by Bonanno and Urpin (\cite{BnU06}) (henceforth paper I) is an attempt to understand the transport of angular momentum by studying the instability arising in differentially rotating compressible flows in the presence of high magnetic field. The magnetic field had nonvanishing radial and azimuthal components. They found a new instability which survived for any value of magnetic field unlike MRI which survives only in the weak field limit. The maximum growth rate of the reported instability was $\sim \Omega$ where $ \Omega$ is the rotation frequency.

This work investigates the magnetohydrodynamic instability in an relatively unexplored region in accretion disk described by $\Omega_e^2<0$, $\Omega_e$ being the epicyclic frequency. 

For brevity, we explain the physical meaning of epicyclic frequency. Square of Epicyclic frequency is defined as $ \Omega_e^2  =  4 \Omega^2 (1 + \frac{s}{2 \Omega} \frac{d \Omega}{d s}$), s being the radial distance from the rotation axis, $\Omega$ being the angular velocity of the fluid element in the flow. The physical meaning of the epicyclic frequency is explained as follows: Orbits of stars in a galaxy are usually not exact circles and always have a certain amount of randomness. If one of the perturbed orbits is observed on a framework rotating with the mean angular velocity around the center of the galaxy, it is known to be a small circle (\cite {Bertin}). The frequency of such small cyclic motions that originate from randomness is called the epicyclic frequency.

Negative epicyclic frequency can be understood as follows: Square of epicyclic frequency is defined as 
\begin{equation}
\Omega_e^2 = s^{-3} \frac {d}{ds}(s^2 \Omega)^2
\end{equation}
s is the radial distance from the rotation axis, $s^2 \Omega$ is the specific angular momentum. An inward increase  in the specific angular momentum (i.e, outward decrease in angular momentum) implies that the square of epicyclic frequency is negative. A fluid element displaced outward under conservation of angular momentum has a larger angular momentum compared with that of the surrounding media. In this case,  the centrifugal force acting on the displaced element is stronger than that acting on the surrounding media, and the element goes further outward (unstable) (\cite{Kato1}). In literature, it is known as a Reyliegh criterion for instablity.

It is well known that in a keplerian disk, the specific angular momentum increases with radius, i.e.,  $\Omega_e^2>0$. According to the Rayleigh criterion, such a flow is stable. Generally in accretion disk, the region with $\Omega_e^2>0$ do not exist but such regions are realized locally in the models by \cite{Honma,Manmoto, Kato}. A recent work by \cite{Gracia} has numerically shown that the transition region between ADAF and outer standard disk is time-dependent having $\Omega_e^2<0$. They modelled the time-evolution of bimodal accretion discs where an outer cooling-dominated disc matches an inner ADAF type flow through a narrow transition region. The behaviour of the instability in such regions is explored.

This paper is orgnized as follows: The section 2 investigates the the growth rate of instability in the new region of parameter space. The section 3 deals with the Results and Discussions. Finally, we conclude by summarizing the new findings.

\section{The instability criteria}
Paper I considered  an axisymmetric differentailly rotating system in the presence of a magnetic field.  A cylinderical coordinate system ($s$, $\varphi$, $z$) with s being the radial distance from the rotation axis was constructed. Unperturbed system was decscibed by ($v_r, v_{\phi}, v_z $) = ($0, s \Omega, 0$). Furthermore, $\Omega$ where $\Omega$ being the angular velocity of the astrophysical flow, was taken to be approximately a function of s alone; i.e., $\Omega = \Omega (s)$.  The isothermal flow for a compressible fluid was described by the following MHD equations

\begin{eqnarray}
\dot{\vec{v}} + (\vec{v} \cdot \nabla) \vec{v} = - \frac{\nabla p}{\rho} 
+ \vec{g} + \frac{1}{4 \pi \rho} (\nabla \times \vec{B}) \times \vec{B}, 
\end{eqnarray}
\begin{equation}
\dot{\rho} + \nabla \cdot (\rho \vec{v}) = 0, 
\end{equation}
\begin{equation}
\dot{p} + \vec{v} \cdot \nabla p + \gamma p \nabla \cdot 
\vec{v} = 0,
\end{equation}
\begin{equation}
\dot{\vec{B}} - \nabla \times (\vec{v} \times \vec{B}) + \eta
\nabla \times (\nabla \times \vec{B}) = 0,
\end{equation}
\begin{equation}
\nabla \cdot \vec{B} = 0. 
\end{equation} 
Where $\rho$ and $\vec{v}$ are the density and fluid velocity, respectively; $p$ is the gas pressure; $\vec{g}$ is gravity; $\vec{B}$ is the magnetic field, $\eta$ is the magnetic diffusivity, and $\gamma$ is the adiabatic index. Magnetic field has non vanishing radial and azimuthal components.

Quasistationary state for a differentially rotating flow with magnetic field was considered. Axisymmetric Eulerian perturbations with space time dependence $\propto \exp ( \sigma t - i \vec{k} \cdot \vec{r})$ where $\vec{k}= (k_{s}, 0, k_{z})$ were introduced in the unperturbed accretion disk and the dispersion relation was obtained for a special case of $\vec{k} . \vec{B} = 0$ after neglecting the Ohmic dissipation in the induction equation.

We take the dispersion relation of paper I for determining the stability of axisymmetric short wavelength perturbations . The dispersion relation (Eq. ($16$) of paper I) is given by the  
\begin{equation}\label{e:eq1}
\sigma^{5} + \sigma^{3} (\omega^{2}_{0} + \Omega^{2}_{e})
+ \sigma^{2} \omega^{3}_{B \Omega} + \sigma \mu \Omega^{2}_{e} 
\omega^{2}_{0} + \mu \Omega^{2}_{e} \omega^{3}_{B \Omega} = 0
\end{equation}
where
\begin{eqnarray}
\Omega^{2}_{e} = 2 \Omega
(2 \Omega + s \frac{d \Omega}{ds}) \;, \;\;
\omega^{2}_{0}= k^{2} ( c^{2}_{s} + c^{2}_{m})\; ,\;\;
\mu = k^{2}_{z} / k^{2}\;,
\nonumber\\
c^{2}_{m} = \frac{B^{2}}{4 \pi \rho} \;, \;\; 
c^{2}_{s} = \frac{\gamma p}{\rho} \;, \;\; \omega^{3}_{B \Omega}=
\frac{k^{2} B_{\varphi} B_{s} s \Omega'}{4 \pi \rho} \;,
\nonumber 
\end{eqnarray}

Equation (\ref{e:eq1}) is a polynomial of degree five, so five non trivial roots exist. We apply Routh-Hurwitz method (\cite{Alek}) (see appendix) to find the regions describing instabilities. This methos has been applied in the field of astrophysics very frequently (\cite{BnU06,BnU07a,Balbus2,Balbus1,UnR05,U03}) Instability exist if any of the conditions written below is satisfied.

\begin{equation}
\mu \Omega^{2}_{e} \omega^{3}_{B \Omega} < 0 \;,\;\;
\omega^{3}_{B \Omega} > 0 \;,\;\;
(\omega^{3}_{B \Omega})^{2}  > 0.
\end{equation}

From the above inequalities, it is interesting to note that the instability will exist for $\Omega_e^2 < 0$ also. We will investigate this region for instability.

Polynomial given by Eq. (\ref{e:eq1}) in $\Omega_e^2<0$ region can be written as

\begin{equation}\label{e:eq3}
\sigma^5 + \sigma^3 (\omega_0^2-|\Omega_e^2|) + \sigma^2 \omega_{BQ}^3  - \sigma \mu |\Omega_e^2| \omega_0^2  -
\mu |\Omega_e^2| \omega_{BQ}^3  = 0
\end{equation}

Let us explore the behavour of dispersion relation for incompressible flow. Since incompressibility corresponds to $\omega_0^2 \xrightarrow{}$ $\infty$, Equation (\ref{e:eq3}) becomes 

\begin{equation}
\sigma \left (\sigma ^2 -\mu |\Omega_e^2| \right)= 0
\end{equation}

It is to be noted that only those perturbations are considered here in which the wavevector is perpendicular to the magnetic field, i.e. $\vec{k} \cdot \vec{B}=0$. It is interesting to note that in the compressible limits, the magnetic field, irrespective of its strength, do not play any role. The hydromagnetic instability cease to exist and the fluid behaves as a Rayleigh unstable hydrodynamic entity. This indicates that in the presence of magnetic field, the compressibility of fluid plays a major role in deciding the occurence of hydromagnetic instability for the above dicussed magenetic field configuration. This behaviour, i.e., the dependence of hydromagnetic instability on the nature of fluid (compressible/ incompressible) can be understood by recalling that the magnetorotational instability is a property of incompressible fluid and it cease to exist for the compressible flow.

When we consider the non magnetic case, the dispersion relation yields Reyliegh unstable modes.

Let us investigate the growth rate of the hydromagnetic instability. To calculate the growth rate of this instability in this unexplored region, $\Omega_e^2$ $<0$, it is convenient to introduce the dimesionless quantities 
\begin{equation}\label{e:eq4}
\Gamma= \frac{\sigma}{|\Omega_{e}|} \;,\;\; \xi = 
\frac{1}{x^{2}} \frac{\omega^{2}_{0}}{|\Omega^{2}_{e}|} \;,\;\; 
\zeta = \frac{1}{x^{2}} \frac{\omega^{3}_{B \Omega}}{|\Omega^{3}_{e}|}
\;, \;\; x=ks
\end{equation}
The polynomial given by Eq. (\ref{e:eq3}) in the case of $\Omega_e^2$ $<0$ can be written as
\begin{equation}\label{e:eq5}
\Gamma^5 + \Gamma^3 (\xi x^2 - 1) + \Gamma^2 \zeta x^2 - \Gamma \mu \xi x^2 - \mu \zeta x^2 =0
\end{equation}

This equation is solved numerically (see \cite{Press} for details) by computing the eigen values of the matrix whose charesteristic polynomial is given by equation (\ref{e:eq3})  for different values of parameter $\mu$, $\xi$ and $\zeta$. We get three real roots and one pair of complex root unlike paper I where two pairs of complex conjugate roots ware obtained.

\begin{figure}
\includegraphics[width = 0.25\textwidth, angle=270]{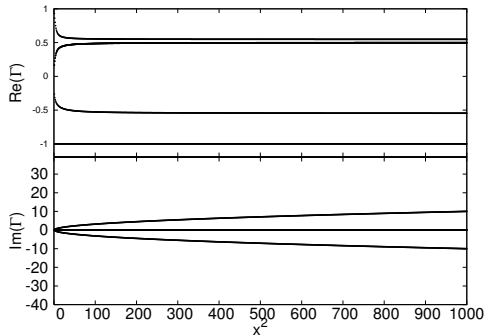}
\caption{The dependence of real and imaginary parts of $\Gamma$ on $x^2$ for $\mu = 0.3$, $\xi = 0.1$, $\zeta = 0.1$}
\end{figure}

Fig. 1, shows the dependence of real and imaginary parts of $\Gamma$ on x for $\mu=0.3$, $\xi = 0.1$, $\zeta =0.1$. Two real roots are obtained when $\mu=0.3$, $\xi=0.1$ and $\zeta=0.1$ whose growth rates are $\sim 0.5 \Omega_e$ but when value of $\xi$ is increased (Fig. 2), growth rate of one of the roots with positive real part deminishes and for any further increase in the value of $\xi$, this root decays very fast and approaches zero but the growth rate of the other root remains around $\sim 0.5 \Omega_e$. For $\xi>>\zeta$, growth rate of one of the roots approaches zero as in paper I but the growth rate of the other root with positive real part remains $\Gamma \sim 0.5 $ and is almost independent of the wavenumber beyond a certain value. It appears that the root in paper I which diminishes with increasing value of $\xi$ is similar to the root which diminishes in our calculations also (it is clear if we compare Fig, 1 of our paper with the Fig. 1 of paper I) but the second surviving positive root in our calculation is possibly a contribution of $\Omega_e^2 < 0$, which was neglected in the paper I. This is qualitatively clear because $\xi>>\zeta$ corresponds to incompressible limit where the reported instability of paper I diminishes and only contribution from $\Omega_e^2 < 0$ remains.

\begin{figure}
\includegraphics[width = 0.25\textwidth, angle=270]{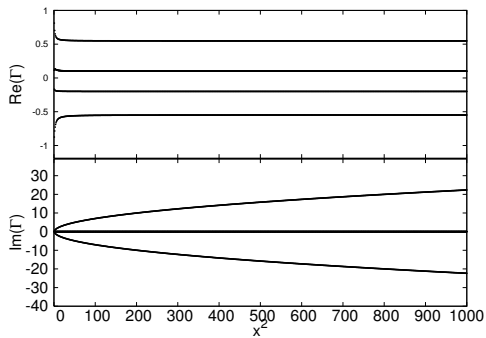}
\caption{The dependence of real and imaginary parts of $\Gamma$ on $x^2$ for $\mu = 0.3$, $\xi = 0.5$, $\zeta = 0.1$}
\end{figure}

But a very interesting result is obtained when we calculate the growth rate of instability for $\zeta > \xi$. Value of $\xi$ was fixed to $0.1$ and the parameter $\zeta$ was varied with $\zeta= (0.5, 1.0,1.5,2.0,3.0$). It is observed that the root in this case is growing faster than the previous cases and this growth is not independent of the wavenumber. Fig. 3, clearly shows this pattern. Interestingly the growth rate of the instability is as high as $5.9 \Omega_e$ for $x^2$ =900 at $\zeta=3.0$. 

\begin{figure}
 \includegraphics[width = 0.25\textwidth, angle=270]{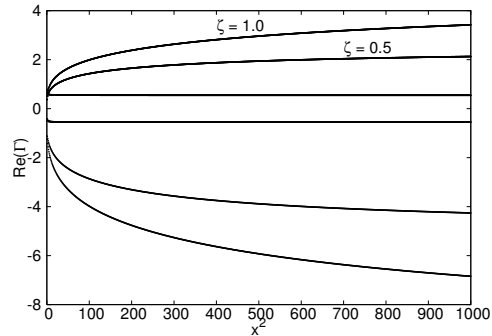}
\caption{The dependence of real parts of $\Gamma$ on $x^2$ for $\mu = 0.3$, $\xi = 0.1$ and $\zeta = 0.5, 1.0$}
\end{figure}

Undoubtedly this is a very strong instability and much more powerful than what was observed in the paper I. The maximum growth rate reported in Paper I was $\sim \Omega_e$ for $x^2 \geq 200$. The instability in our case is very important on two counts, i) Growth rate is very high ($\sim$ $5.9 \Omega_e$) for $x^2 =900$ at $\zeta =3$) ii) unlike the previous report, it depends on the wavelength of perturbation.

\begin{figure}
\includegraphics[width = 0.25\textwidth, angle=270]{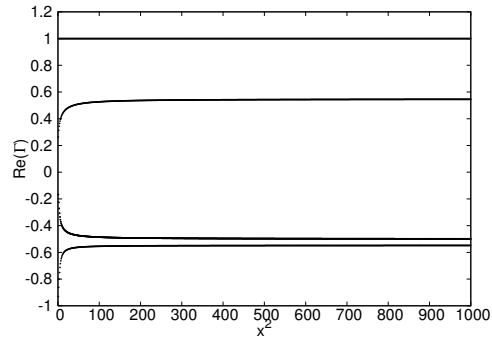}
\caption{The dependence of real $\Gamma$ on $x^2$ for $\mu = 0.3$, $\xi = 0.1$ and $\zeta = -0.1$}
\end{figure}

In Fig.4, we plot the dependence of Re ($\Gamma$) on x for $\mu = 0.3$, $\xi = 0.1$ and $\zeta = -0.1$. It is very clear that the change of sign of $\zeta$ reverses the behaviour of roots (see Fig. 1, and Fig. 4.). We do not plot the growth of imaginary part of root as we are interested in the instability. 

\section{Results and Discussion} 
We revisit the problem of hydromagnetic instability in differentially rotating compressible flows and analyse the instability. The behaviour of hydromagnetic instability is investigated in the region described by $\Omega_e^2<0$. Though $\Omega_e^2<0$ is generally not satisfied in accretion disk but there are possibilities that $\Omega_e^2<0$ is locally satisfied (\cite{Gracia,Manmoto,Kato,Honma}) in the accretion disks. The present work will be applicable in those regions (\cite{Kato2}).

We discover a range of parameter space described by $\xi$, $\zeta$ and $\mu$ where the instability grows very fast, much faster than the reported growth rate of the instability in paper I. We have shown the growth rate of the instability in Fig. 3, for two representative values of $\zeta$ for fixed $\xi$. Growth rate  is $\sim 1.8 \Omega_e$, $ \sim 2.0 \Omega_e$ and $ \sim 2.1 \Omega_e$ at $x^2 = 300$, $600$ and $900$ respectively for $\zeta= 0.5$ and the growth rate is $\sim 2.6 \Omega_e$ and $ \sim 3.0 \Omega_e$ and $ \sim 3.4 \Omega_e$ respectively for same values of $x^2$ for $\zeta= 1.0$. The growth rate increases further with increasing value of $\zeta$ for fixed values of $\xi$. It is surprising to observe that the instability grows very fast and the growth rate, unlike in paper I is not independent of wavelength of perturbation, Infact, the instability grows with the increasing value of $x^2$.

In the case of $\xi >> \zeta$, which corresponds to the incompressible limit, one of the roots diminishes as was in paper I but the second root with postive real part survives with growth rate $ 0.547732  \Omega_e$. Since incompressibility corresponds to $\omega_0^2 \xrightarrow{}$ $\infty$, Equation (\ref{e:eq3}) becomes 
\begin{equation}
\sigma \left (\sigma ^2 -\mu |\Omega_e^2| \right)= 0
\end{equation}
It is imperative to see that in the case of incompressible flow, Eq.(\ref{e:eq3}) reduces to hydrodynamic relation. Since $\mu = 0.3$, growth rate comes to $\sqrt{\mu} \Omega_e$, i.e., $.0547732$. Hence, it is clear that the root which survives in the incompressible limit corresponds to the root satisfying Rayleigh criterion of instability in hydrodynamics.

\section{Conslusion}
This paper considers a generalization of the well known Rayleigh instability for the case of compressible magnetic flows. The growth rate of the instability in the new region $\Omega_e^2 < 0$ is rather large. Moreover, the instability in our case also depends on the wave number. High growth rate and dependence on wavenumber makes this instability very interesting. Since the instability exists even in the high magnetic field where MRI is supressed, it will be interesting to study the effect of this instability in angular momentum transport.

\section*{Acknowledgments}
The author thanks S. Kato for useful comments. The author expresses his sincere thanks to the referee for useful suggestions. The author also thanks A.K.Mitra whose suggestions lead to a considerable improvement in the presentaion of the manuscript.
\appendix

\section{Hurwitz method}
\noindent Let us consider an $5^{th}$ order polynomial P(x)
\begin{equation}\label{e:app1}
P(x)= a_5 x^5 + a_4 x^4 + a_3 x^3 +a_2 x^2 +..... +a_0
\end{equation}
Then Hurwitz theorem states that the above polynomial will be unstable if any of the following inequalities is satified (\cite{UnR05}).
\begin{eqnarray}
\lefteqn{a_{0} <0 \;,}
\nonumber \\
\lefteqn{A_{1} \equiv a_{4} a_{3} - a_{2} < 0 \;, }
\nonumber \\
\lefteqn{A_{2} \equiv a_{2} (a_{4} a_{3} -a_{2}) - a_{4}(a_{4} a_{1} 
- a_{0}) <0 \;,}
\nonumber \\
\lefteqn{A_{3} \equiv (a_{4} a_{1} -a_{0}) [a_{2} (a_{4} a_{3} 
- a_{2}) - a_{4}
(a_{4} a_{1} - a_{0})] - }
\nonumber \\
&& \quad \quad \quad \quad \quad \quad - a_{0} (a_{4} a_{3} 
- a_{2})^{2} < 0 \;,
\label{13}
\end{eqnarray}

\bibliographystyle{mn2e}
\bibliography{MradulRevised}

\label{lastpage}

\end{document}